\documentclass[10pt]{iopart}
\usepackage{iopams}  
\newcommand{\eqref}[1]{\eref{#1}}

\usepackage{graphicx}

\usepackage[utf8]{inputenc}
\usepackage[T1]{fontenc}
\usepackage{amssymb}
\usepackage{amsfonts}

\usepackage{bm}
\usepackage{color}
\usepackage{graphicx}
\usepackage{float}

\usepackage{courier}
\usepackage{txfonts}

\newcommand{\unit}[1]{\ensuremath{\bm{\widehat{\mathrm{#1}}}}}
\usepackage{bm}
\usepackage{color}
\usepackage{graphicx}

\newcommand{\sub}[1]{\ensuremath{_{\rm #1}}}
\newcommand{\rd}{\ensuremath{\mathrm{d}}}

\usepackage{ae}
\usepackage{mathrsfs}
\usepackage{units}
\usepackage{todonotes}% [disable]
\usepackage[T1]{fontenc}
\usepackage{bm}
\usepackage{color}
\usepackage{graphicx}
\usepackage[breaklinks=true]{hyperref}
\hypersetup{
  unicode=false,          % non-Latin characters in Acrobat’s bookmarks
  pdftitle={Runaway avalanche generation during massive material injection},    % title
  pdfauthor={L Hesslow},     % author
  pdfnewwindow=true,      % links in new PDF window
  colorlinks=true,        % false: boxed links; true: colored links
  linkcolor=blue,         % color of internal links (change box color with linkbordercolor)
  citecolor=blue,         % color of links to bibl
  filecolor=blue,         % color of file links
  urlcolor=blue           % color of external links
}

\usepackage{enumerate}

\newcommand{\nofrac}[2]{#1/#2}

\newcommand{\Eceff}{\ensuremath{E_{\rm c}^{\rm eff}}}

\newcommand{\Ec}{E_{\rm c}}

\newcommand{\netot}{n_{\rm e}^{\rm tot}}
\newcommand{\nef}{n_{\rm e}}

\newcommand{\nuSBar}{\bar{\nu}_{\rm s}}
\newcommand{\nuDBar}{\bar{\nu}_{\rm D}}

\newcommand{\lnLc}{\ln\Lambda_{\rm c}}

\begin{document}

%\title{Influence of massive material injection on runaway avalanching}
\title[Runaway avalanche generation during massive material injection]{Influence of massive material injection on avalanche runaway generation during tokamak disruptions}
%Enhancement of the avalanche multiplication factor due to massive material injection and implications for runaway mitigation
\author{L~Hesslow, O~Embr\'eus, O~Vallhagen and T~F\"ul\"op}
\address{Department of Physics, Chalmers University of Technology, 41296 Gothenburg, Sweden}
\ead{hesslow@chalmers.se}

\vspace{10pt}
\begin{indented}
\item[]\today
\end{indented}

\begin{abstract}
In high-current tokamak devices such as ITER, a runaway avalanche can cause a large amplification of a seed electron population. We show that disruption mitigation by impurity injection may significantly increase the runaway avalanche growth rate in such devices. This effect originates from the increased number of target electrons available for the avalanche process in weakly ionized plasmas, which is only partially compensated by the increased friction force on fast electrons. We derive an expression for the avalanche growth rate in partially ionized plasmas and investigate the effects of impurity injection on the avalanche multiplication factor and on the final runaway current for ITER-like parameters. For impurity densities relevant for disruption mitigation, the maximum amplification of a runaway seed can be increased by tens of orders of magnitude compared to previous predictions. This motivates careful studies to determine the required densities and impurity species to obtain tolerable current quench parameters, as well as more detailed modeling of the runaway dynamics including transport effects.

\end{abstract}


\submitto{\NF}
%\maketitle
\ioptwocol

{\it Introduction.}---Disruptions and their risk of creating a runaway-electron beam are considered a serious threat to the successful operation of large tokamaks. 
A robust and thoroughly validated disruption mitigation scheme is therefore critical, and remains one of the outstanding challenges for ITER~\cite{Lehnen2015}. 
Successful disruption mitigation must both limit the forces and thermal loads from the disruption, and prevent runaway beam formation insofar as this is possible.  
In order to limit the forces from halo and eddy currents on the vacuum vessel, the current quench duration must be kept within a time interval of approximately \unit[50-150]{ms}, which is typically achieved by injection of high-Z impurities~\cite{HollmannDMS}. 

The induced electric field in the current quench after impurity injection has the potential to create a large runaway beam through the avalanche effect, whereby runaway electrons multiply by large-angle collisions with cold electrons~\cite{sokolov1979multiplication, jayakumar1993,RosenbluthPutvinski1997}. The maximum multiplication of a runaway seed is exponentially sensitive to the plasma current, meaning that the runaway avalanche can become problematic in reactor-scale tokamaks~\cite{RosenbluthPutvinski1997}. The avalanche multiplication factor is therefore a key parameter 
for runaway generation in high-current devices; in particular, it determines the maximum plasma current above which a small seed can be multiplied into a dangerous runaway current~\cite{Boozer2019}.

Apart from its effect on the thermal and current quench dynamics, impurity injection has a direct effect on runaway dynamics.
Early calculations indicated that an increased average ion charge would reduce the avalanche growth rate~\cite{RosenbluthPutvinski1997}, in which case a disruption mitigation system based on massive material injection would also act to suppress runaway generation during the current quench. %~\cite{Parks1999}. 
However, even though the collision rates are increased significantly in the presence of partially ionized high-$Z$ impurities~\cite{Hesslow,Ecrit}, 
recent work has indicated that the avalanche growth rate is in fact increased in the presence of strong electric fields and for certain plasma compositions~\cite{HesslowJPP,McDevitt2019_avalanche}. This is because
the increased dissipation is counteracted by an even larger increase in the number of target electrons available for runaway multiplication by large-angle collisions. 
Consequently, the effect of partially ionized impurities on the avalanche growth rate must  be accounted for in runaway modeling.

In this Letter, we investigate the effect of massive gas injection on avalanche runaway generation. 
We derive an expression for the avalanche growth rate,
which we use to determine the avalanche multiplication during a current quench. In order to isolate the avalanche dynamics, we consider
an idealized scenario where a thermal quench has produced a cold  partially ionized plasma containing a seed runaway population. Under these circumstances, we analyze the subsequent avalanche multiplication of such a surviving seed.

{\it Runaway avalanche multiplication in tokamaks.}---The avalanche runaway dynamics are governed by the growth-rate equation 
\begin{equation}
\frac{\partial n\sub{RE}}{\partial t} = \Gamma n\sub{RE},
\label{eq:general growth rate}
\end{equation}
where $n\sub{RE}$ is the runaway electron density, $\Gamma$ denotes the instantaneous avalanche growth rate, and we have neglected radial transport and toroidal effects of runaways, which have been analyzed recently in \cite{McDevitt2019_avalanche,McDevitt2019_transport}. Ever since the first identification of runaway avalanche multiplication in plasmas by Sokolov~\cite{sokolov1979multiplication}, it has been predicted that for large electric fields---far exceeding the avalanche threshold---the avalanche growth rate in a plasma is approximately proportional to the electric field component along the magnetic field, $\Gamma \,{\approx}\,\Gamma_0 E_\parallel$. In a tokamak, especially near the magnetic axis where the bulk of runaway generation often occurs~\cite{Smith2006GO}, the magnetic field is nearly aligned with the toroidal direction, and the growth rate equation takes the form
\begin{equation}
\frac{\partial}{\partial t}\ln n\sub{RE} \approx \Gamma_0 E_\varphi \equiv -\Gamma_0 \frac{\partial A_\varphi}{\partial t}.
\end{equation}
Here, the toroidal component of the electromagnetic vector potential is given by $A_\varphi{\,=\,}-\psi_p/2\pi R$, with $\psi_p$ the poloidal flux and $R$ the major radius of the tokamak. Assuming an initial runaway seed population with density $n\sub{seed}$ at the end of the thermal quench, we may write the resulting runaway density as
\begin{equation}
n\sub{RE} = n\sub{seed} \exp(N\sub{ava}),
\end{equation}
where $N\sub{ava}$ is the logarithm of the avalanche multiplication factor. When $\Gamma_0$ does not change appreciably in time, $N\sub{ava}$ becomes independent of the details of the time evolution of the electric field, and is completely determined by the net change in the local poloidal flux:
\begin{equation}
N\sub{ava} = \frac{\Gamma_0\Delta \psi_p}{2\pi R_0}.
\end{equation}
where $R_0$ is the major radius of the magnetic axis. 
In ITER, the central poloidal flux may be of the order of $\psi_p{\,\approx\,}70$-100\,Vs~\cite{Boozer2015}, which implies maximum multiplication factors in the range of $\exp(N\sub{ava}){\,\approx\,} 10^{13}$-$10^{17}$ near the core, using the Rosenbluth--Putvinski expression for the growth rate $\Gamma_0{\,=\,}e/(\ln\Lambda m_{\rm e} c\sqrt{5+Z\sub{eff}})$~\cite{RosenbluthPutvinski1997} with $\ln\Lambda {\,=\,}15$ and a plasma effective charge $Z\sub{eff} \,{\equiv}\,\sum_j n_j Z_j^2/n_{\rm e}\,{=}\, 1$.
For a fixed runaway seed it is thus the product of avalanche growth rate and poloidal flux change, $\Gamma_0 \Delta \psi_p$, that determines whether an unacceptably large runaway beam will form. 

In the presence of partially ionized impurities, the avalanche growth rate will no longer be directly proportional to the electric field, as assumed above, and the multiplication factor will instead become sensitive to the details of the $E$-field evolution. In the large-aspect-ratio limit with circular concentric flux surfaces, the parallel current and electric field satisfy
\begin{equation}
\mu_0\frac{\partial j}{\partial t} = \frac{1}{r}\frac{\partial}{\partial r}\left(r\frac{\partial E_\parallel}{\partial r}\right),
\label{eq:Ediff}
\end{equation}
which has been widely used in disruption runaway modeling~\cite{Smith2006GO,MartinSolis2017,Papp_2013,breizman2014}. Here, $r$ denotes the distance from the magnetic axis, and we assume there is a conducting wall at the minor radius $r{\,=\,}a$, where $E(a)=0$.
An approximation of the avalanche multiplication factor can be obtained in the trace-runaway limit, $j = \sigma E_\parallel$, and the conductivity $\sigma$ is constant. An eigensolution of~\eqref{eq:Ediff} is then given by
\begin{equation}
    j(r,t=0) = \frac{x_1}{2}\frac{I_0}{\pi a^2}\frac{J_0(x_1 r/a)}{J_1(x_1)},
\end{equation}
where $I_0$ is the total initial plasma current, $J_n$ denotes the $n$'th Bessel function of the first kind and $x_1 \,{\approx}\,2.4$ is the first zero of $J_0$. In this case, the electric-field evolution is given by $\sigma E(r,\,t) = j(r,0) \exp(-t/t\sub{CQ})$, where the 
current quench e-folding time is 
\begin{equation}
t\sub{CQ} = \frac{\sigma \mu_0 a^2}{x_1^2},
\end{equation}
which should be approximately in the range of \unit[22-66]{ms}~\cite{MartinSolis2017} to be consistent with a total current quench duration of \unit[50-150]{ms}.
This allows us to explicitly integrate the runaway growth rate equation~\eqref{eq:general growth rate} to give the maximum avalanche multiplication factor through
\begin{equation}
N\sub{ava}(r) \approx t\sub{CQ}\int_{\Eceff(r)}^{E\sub{initial}(r)} \frac{\Gamma(E,r)}{E}\rd E,
\label{eq:GammaOverE}
\end{equation}
where $\sigma E\sub{initial} = j(r,0)$, and the effective critical electric field $\Eceff$ is defined by $\Gamma(\Eceff,r)\equiv 0$. This result allows us to qualitatively assess how a change in the growth rate $\Gamma$ results in a modified avalanche amplification of the runaway population during a current quench in a disrupting tokamak plasma.

\emph{Avalanche growth rate in partially ionized plas\-mas.}---To determine the avalanche growth rate in a partially ionized plasma, we consider the kinetic equation in a uniform and magnetized system: 
\begin{eqnarray}
 \tau_c\frac{\partial {f}}{\partial t} - 
  \frac{E_\parallel}{E_{\rm c}} \left(\xi\frac{\partial f}{\partial p}+\frac{1-\xi^2}{p}\frac{\partial f}{\partial \xi} \right) 
  = 
\nonumber  \\ \quad
  \frac{1}{p^2}\frac{\partial}{\partial p} 
\left( \nuSBar \gamma^2{f} \right)
+
\frac{1}{2}\nuDBar\frac{\gamma}{p^3}\frac{\partial}{\partial \xi}\left((1-\xi^2)\frac{\partial {f}}{\partial \xi}\right) 
+ \tau_c C_{\rm ava}\,,
\label{eq:FP}
\end{eqnarray}
where $f$ is the electron distribution function, $p\,{=}\,\gamma v/c$ is the normalized momentum, $E_\parallel$ is the component of the electric field which is parallel to the magnetic field $\bi{B}$ 
and $\xi\,{=}\,\bi{p}\,{\cdot}\,\bi{B}/(p B)$ is the cosine of the velocity pitch-angle. 
The large-angle collision operator, which gives rise to the runaway avalanche, is denoted $C_{\rm ava}$, and it is assumed that the critical runaway energy far exceeds the ionization energy such that this source is proportional to the total (free plus bound) electron density $\netot$. A relaxation of this assumption, which more carefully accounts for the binding energy of the target electrons, has recently been proposed~\cite{McDevitt2019_avalanche}. This would decrease the avalanche growth rate at strong electric fields ($E\,{\gtrsim}\,4000 \Ec$ in argon-dominated plasmas), but the effect tends to be small for the majority of runaway generation during tokamak disruptions. 
The normalized deflection and slowing-down frequencies $\nuDBar(p)$ and $\nuSBar(p)$ account for the effect of partial screening as described in (2.22) and (2.31) of~\cite{HesslowJPP} and reduce to the completely screened limit (of a fully ionized plasma with the same net charge) $\nuDBar\,{\rightarrow}\,1\,{+}\,Z\sub{eff}$ and $\nuSBar\,{\rightarrow}\,1$, if the energy variation in the Coulomb logarithm is ignored.
 Letting $n_{\rm e}$ denote the density of free electrons, the relativistic collision time is $\tau_{\rm c}\,{=}\,\nofrac{4 \pi \epsilon_0^2 m_{\rm e}^2 c^3}{(n_{\rm e} e^4 \lnLc)}$,  with the relativistic Coulomb logarithm 
$ \lnLc \,{\approx}\,14.6+0.5 \ln (T_{\rm eV}/n_{\rm e20})$, where $T_{\rm eV}$ is the electron temperature in electronvolt and $n_{\rm e20}$ is the density of the background electrons in units of $\unit[10^{20}]{m^{-3}}$~\cite{HesslowJPP}. The (Connor--Hastie) critical electric field is $E_{\rm c}\,{=}\,m_{\rm e} c/(e \tau_{\rm c})$, which may be substantially lower than the effective critical electric field \Eceff,  as the latter accounts for radiation as well as collisions with partially ionized impurities. 

To determine the avalanche growth rate, we consider the following three regions in electric-field strength and pitch-angle scattering rate, similarly to Rosenbluth and Putvinski~\cite{RosenbluthPutvinski1997}: 
\vspace{-3mm}\paragraph{(i)  $\nuDBar \,{\gg}\, E_\parallel /\Ec \,{\gg}\, 1$: }
Here, we adopt an ordering in a small parameter $\delta$ according to
\refstepcounter{equation}
$$
  \frac{\nuDBar}{\netot/\nef} \sim \delta^{-2}, \qquad \frac{E_\parallel}{\Eceff} \sim \delta^{-1},
  \eqno{(\theequation{\mathit{a},\mathit{b}})}
  \label{eq:ordering_def}
$$
Solving~\eqref{eq:FP} to order $\delta^0$ gives an expression for the growth rate in terms of a weighted integral over the collision frequencies. If 
$
\nofrac{E_\parallel^2}{\Ec^2} \,{\gg}\,3 \nuSBar \nuDBar,
$
which is consistent with the ordering in~\eqref{eq:ordering_def}, the resulting growth rate is approximately given by 
\begin{equation}
\Gamma \approx \frac{e}{m_{\rm e} c \lnLc} \frac{\netot}{\nef}
\frac{E_\parallel}{\sqrt{\nuSBar(p_\star) \nuDBar(p_\star)}}\, ,\label{eq:region2Sol}
\end{equation}
where
$
p_\star  
=  \nofrac{\sqrt[4]{\nuSBar(p_\star) \nuDBar(p_\star)}}{ \sqrt{E_\parallel/\Ec } }
$
takes the form of an effective critical momentum for runaway acceleration. Being defined only implicitly through this relation, it must in general be evaluated numerically.
\vspace{-3mm}\paragraph{(ii) $E_\parallel/\Eceff \,{\approx}\,1$:}
We demand that the avalanche growth rate vanishes at the effective critical electric field. Therefore, for weak electric fields, 
\begin{equation}
\Gamma \propto (E_\parallel-\Eceff).
\label{eq:region3Sol}
\end{equation} 
We use the expression for \Eceff\ derived in~\cite{Ecrit}~\footnote{A numerical implementation of \Eceff\ is available at \url{https://github.com/hesslow/Eceff}.}, which accounts for screening effects and radiation reaction.
\vspace{-3mm}\paragraph{(iii) $E_\parallel/\Eceff \,{\gg}\,  1, \nuDBar  = 0$:}
In order to describe plasmas with low impurity content, we take the completely screened approximation of the slowing-down frequency $\nuSBar \,{\approx}\,1$, which gives a growth rate of the same form as in ~\cite{RosenbluthPutvinski1997}; 
\begin{equation}
\Gamma  \approx 
\frac{1}{2}  \frac{e}{m_{\rm e} c \lnLc}\frac{ \netot}{\nef} E_\parallel.
\label{eq:region1Sol}
\end{equation}

The expressions~\eqref{eq:region2Sol}, \eqref{eq:region3Sol} and~\eqref{eq:region1Sol} can be combined into an interpolated formula similarly to Rosenbluth--Putvinski~\cite{RosenbluthPutvinski1997}, valid for $E_\parallel\gtrsim \Eceff$\footnote{In order to produce a well-behaved formula also for $E_\parallel<\Eceff$, which may be used to approximately describe runaway decay at near-critical electric fields, one may replace $p_\star = p_\star(\Eceff)$ for $E_\parallel<\Eceff$.}, giving
\begin{equation}
\Gamma =    \frac{e}{m_{\rm e} c \lnLc} \frac{\netot}{\nef} \frac{E_\parallel-\Eceff}{\sqrt{4+
\nuSBar(p_\star) \nuDBar(p_\star) 
}}.
\label{eq:GammaInterp}
\end{equation}
We note that, for weakly ionized plasmas dominated by heavy impurities, the behavior of $\bar\nu\sub{s}$ and $\bar\nu\sub{D}$ are such that we observe approximately the $\Gamma \propto E_\parallel^{3/2}$ scaling that has been predicted for runaway breakdown in air~\cite{Gurevich2001}, which involves similar physics. 
In a fully ionized plasma, \eqref{eq:GammaInterp} reduces to the simplified version of the growth rate given by Rosenbluth--Putvinski~\cite{RosenbluthPutvinski1997}:
\begin{equation}
\Gamma \rightarrow \frac{e}{m_{\rm e} c \lnLc} \frac{E_\parallel-\Ec}{\sqrt{5+Z_{\rm eff}
}}, \quad \nuSBar\nuDBar\rightarrow 1+Z_{\rm eff}.
\label{eq:RP formula}
\end{equation}

\begin{figure}
\centering
\includegraphics[width=(0.3\linewidth+0.3\textwidth),trim=25mm 0 30mm 0]{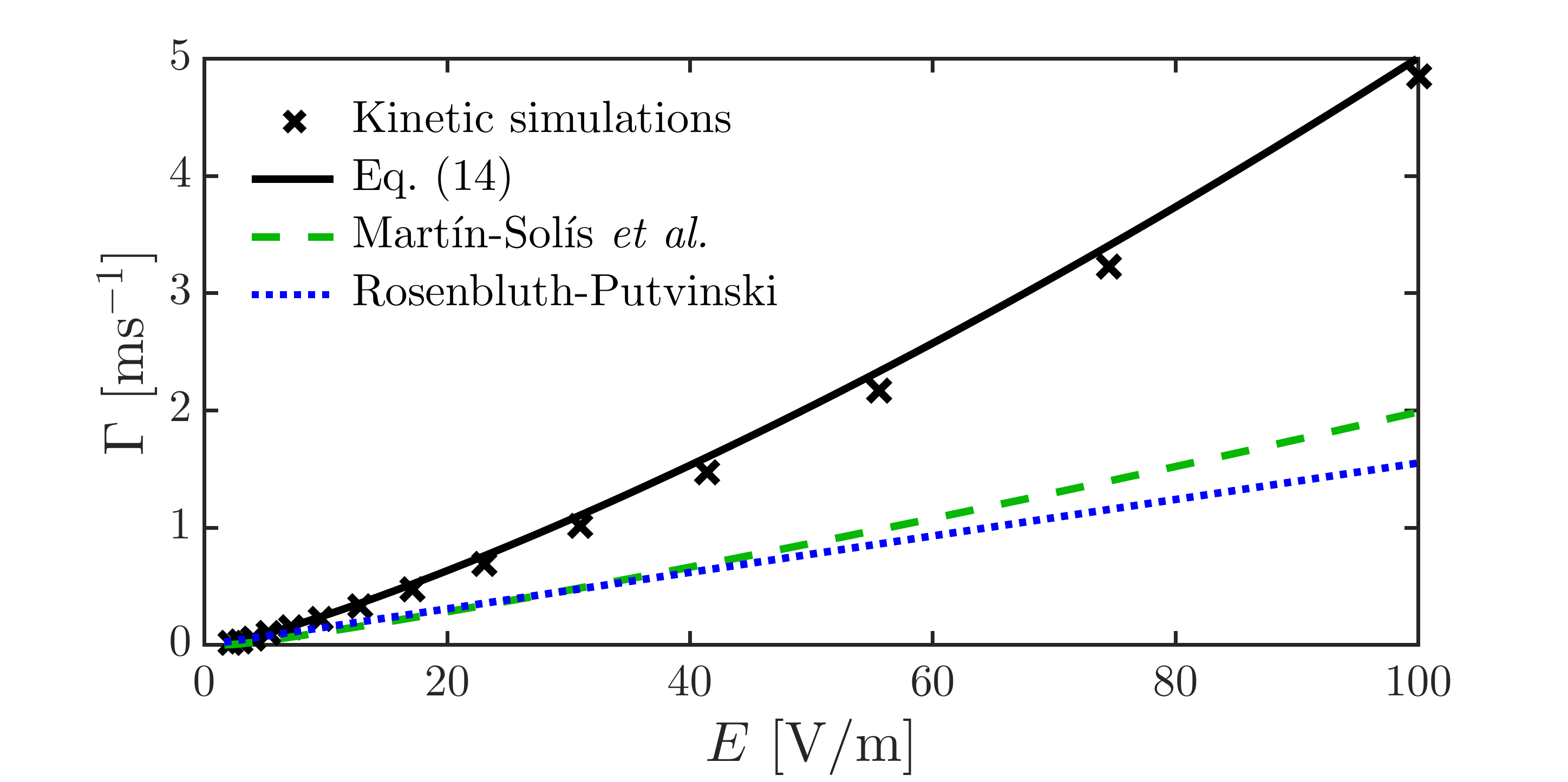}
\caption{\label{fig:BenchmarkCODE} Comparison of the expression~\eqref{eq:GammaInterp} (black, solid) to kinetic simulations using CODE (black, crosses); the model employed in~\cite{MartinSolis2017} (green, dashed); and the Rosenbluth--Putvinski result~\eqref{eq:RP formula} (blue, dotted). The plasma consists of fully ionized deuterium, $n\sub{D} = \unit[10^{20}]{m^{-3}}$, and singly ionized argon with $n_Z = n\sub{D}$, at a temperature $T = 10\,$eV and $\Eceff \,{\approx}\,1.7\,$V/m. 
}    
\end{figure}

We validate the avalanche growth rate formula~\eqref{eq:GammaInterp} by comparing to simulations using a Fokker--Planck solver, \textsc{code}~\cite{CODE,Stahl2016}, which is equipped with a field-particle Boltzmann operator for avalanche generation to model $C\sub{ava}$~\cite{olaknockon2018}.
Our expression gives accurate predictions from $E_\parallel \,{\approx}\,\Eceff$ to strong electric fields and from trace impurity contents to impurity densities $n_Z \,{\gg}\,n\sub{D}$. The maximum deviation between the simulated results and~\eqref{eq:GammaInterp} was of the order of \unit[20]{\%} for all considered impurity ions (Ar$^{0}$, Ar$^{+}$, Ar$^{2+}$, Ar$^{5+},$ Ar$^{7+}$, Ar$^{10+}$, Ne$^{0}$, Ne$^{1+}$ and Ne$^{5+}$). As an example, figure~\ref{fig:BenchmarkCODE} shows the avalanche growth rate for Ar$^+$, including comparisons with the Rosenbluth--Putvinski result~\eqref{eq:RP formula} and the model employed in~\cite{MartinSolis2017}.
Note that our results for the avalanche growth rate differ substantially from previously used results. Compared to our results, \cite{MartinSolis2017} for example  obtains approximately half our growth rate at high impurity densities as shown in figure~\ref{fig:BenchmarkCODE}, but up to twice our predicted values at argon densities much lower than the deuterium density.
This is likely due to their use of both a simplified model for the screening effects and a test-particle model of the avalanche dynamics.

\begin{figure}[!htb]
 \centering
   \includegraphics[width=(\linewidth)]{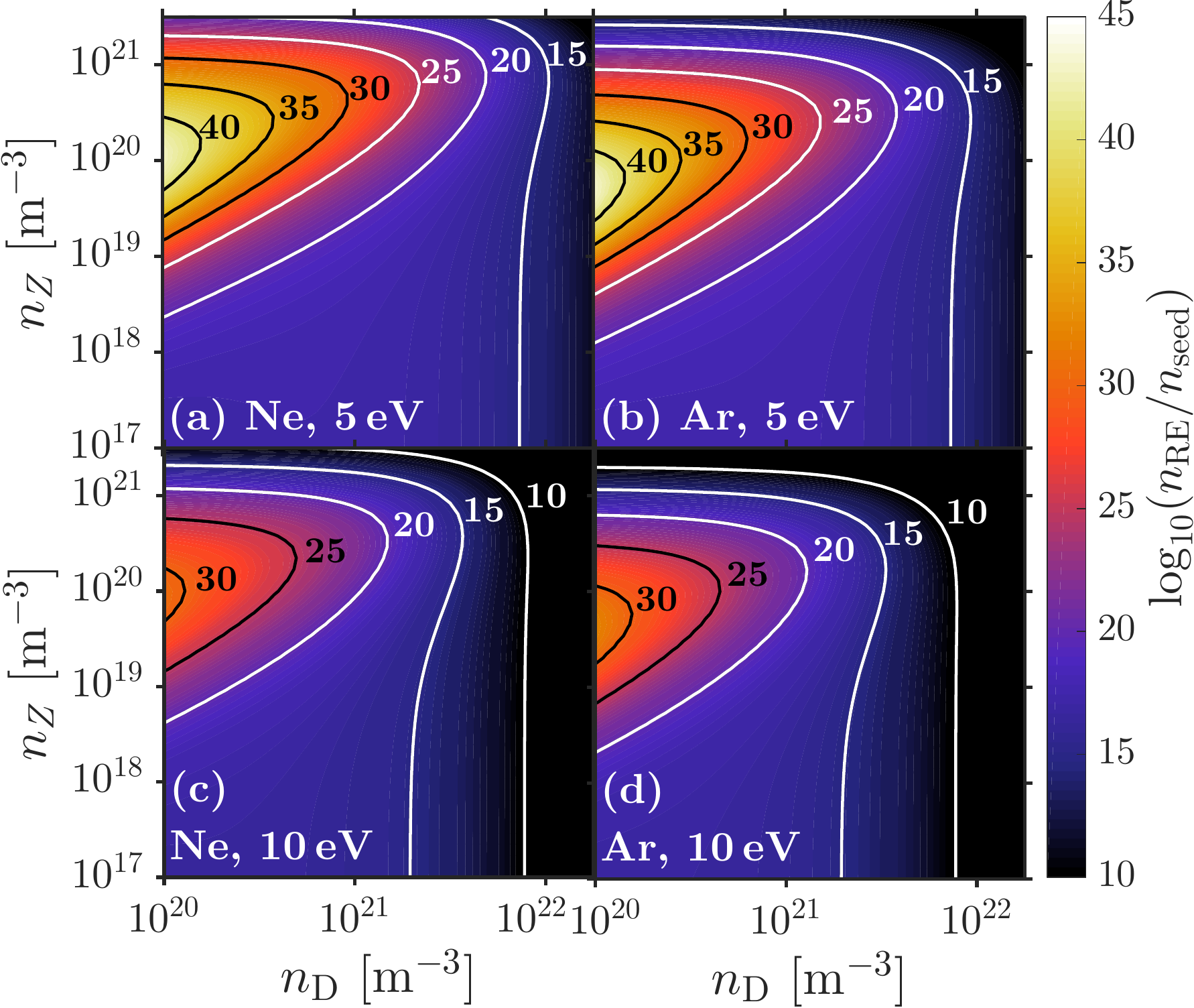}
 \caption{\label{fig:mult1}The decimal logarithm of the avalanche multiplication factor given by~\eqref{eq:GammaOverE}
 as a function of deuterium and impurity density for (a,c) neon and (b,d) argon. The impurity charge states were chosen from a collisional-radiative equilibrium at \unit[5]{eV} (top panels) and \unit[10]{eV} (bottom panels).}
\end{figure}

{\it Effect of massive material injection on the avalanche multiplication factor.}---Having derived an expression for the avalanche growth rate~\eqref{eq:GammaInterp}, we can now address the effect of different plasma compositions on the avalanche multiplication factor using~\eqref{eq:GammaOverE}. Figure~\ref{fig:mult1} shows the logarithm of the on-axis avalanche multiplication factor $n\sub{RE}/n\sub{seed}$, as a function of density of deuterium and Ne or Ar impurities, in an ITER-like scenario with minor radius $a\,{=}\,\unit[2]{m}$
and plasma current $I_0\,{=}\,\unit[15]{MA}$, evaluated using~\eqref{eq:GammaOverE}. In this case, the Rosenbluth--Putvinski growth rate~\eqref{eq:RP formula} predicts a multiplication factor of approximately $10^{14}$-$10^{16}$; our model predicts this number to increase dramatically by accounting more carefully for the presence of partially ionized impurities. For example, as shown in figure~\ref{fig:mult1}, the avalanche multiplication factor at neon  or argon densities around $n_Z\,{=}\,\unit[10^{20}]{m^{-3}}$ is tens of orders of magnitude higher than the Rosenbluth--Putvinski prediction, without deuterium injection.  Conversely, with significant deuterium injection the Rosenbluth--Putvinski prediction is recovered. We note that deuterium densities of $n_D\,{\gtrsim}\,\unit[10^{22}]{m^{-3}}$ would be needed to suppress the runaway avalanche.

As of yet, there is considerable uncertainty in the literature regarding the required amount of injected impurities to obtain a certain current quench time; for example, Mart{\'{\i}}n-Sol{\'{\i}}s {\it et al.}\ report that impurity density much lower than the background plasma density are sufficient~\cite{MartinSolis2017}, whereas Feh\'er {\it et al.}\ report required impurity densities of the order of the plasma density~\cite{Feher}. To reflect these uncertainties, we here assume a fixed background temperature of $T\,{=}\,\unit[5]{eV}$ or \unit[10]{eV}, and illustrate how different impurity densities modify the avalanche multiplication. Accordingly, we  determine the impurity charge states by assuming collisional-radiative equilibrium for the impurities using rate coefficients from the ADAS database~\cite{ADAS}; at \unit[5]{eV}, this gives an effective net charge $Z_{{\rm eff}}\,{=}\,\sum_j n_j Z_0^2/\sum_j n_j Z_0$ of 2.0 for neon and 2.7 for argon (in the absence of deuterium), while the corresponding values at \unit[10]{eV} are 3.0 for neon and 3.8 for argon. 

With partial screening, the current quench time affects the avalanche multiplication factor through the maximum induced electric field; as shown in the bottom panels of figure~\ref{fig:mult1}, massive material injection has a greater effect on the avalanche multiplication factor at lower temperatures, corresponding to shorter current quench times (here, $t\sub{CQ}$ was in the range \unit[12-27]{ms} at \unit[5]{eV}, and \unit[20-68]{ms} at \unit[10]{eV}). 
Notably, in previous models where $\Gamma{\, \propto\,} E_\parallel$, the temperature and density dependence (and thus the current quench time) cancels out in~\eqref{eq:GammaOverE}, except for a weak dependence through the Coulomb logarithm. In such idealized models, it is therefore only the dependence of the avalanche growth rate on $Z\sub{eff}$ that significantly affects the avalanche multiplication factor.  

Impurity injection increases the runaway current even in non-trace scenarios, where the runaway current contributes to the electric-field evolution. 
To illustrate this effect, we used the numerical tool \textsc{go}~\cite{Feher,Papp_2013} to solve~\eqref{eq:Ediff}, accounting for a finite runaway electron current through $j{\, =\, }\sigma E_\parallel{\,+\, }e cn\sub{RE}$, where we let the runaway current evolve according to~\eqref{eq:general growth rate} and~\eqref{eq:GammaInterp}.
We use an ITER-like scenario described by Fehér \emph{et al}.~\cite{Feher}, with an initial plasma current of \unit[15]{MA}, temperature $T{\,=\,}\unit[5]{eV}$, magnetic field \unit[5.3]{T}, major radius \unit[6.2]{m}, minor radius \unit[2]{m}, radius of the conducting wall \unit[2.15]{m} and a radial current profile such that the on-axis poloidal flux is $\psi_p(0){\,\approx\,}\unit[90]{Vs}$. 
For example, at an impurity density of $n_Z {\,=\,} n\sub{D}{\,=\,} \unit[10^{20}]{m^{-3}}$,
a small constant runaway seed of $n\sub{seed}{\,=\,}\unit[10^3]{m^{-3}}$   produces a relatively modest final runaway current with the completely screened Rosenbluth--Putvinski growth rate~\eqref{eq:RP formula}: \unit[1.1]{MA} for neon versus \unit[0.9]{MA} for argon. In contrast, the resulting runaway current is substantial when partial screening is accounted for: 
% Ne:  med screening max(I_run)=6.99 MA, full screening 1.14 MA
% Ar:  med screening max(I_run)=7.05 MA, full screening 0.91 MA
\begin{equation*}
I\sub{RE} = \unit[7.0]{MA} \,({\rm Ne}), \quad I\sub{RE} = \unit[7.1]{MA} \,({\rm Ar}).
\end{equation*}
% Ne:  med screening max(I_run)=2.28 MA, full screening 1.64 MA
% Ar:  med screening max(I_run)=2.49 MA, full screening 1.63 MA
For lower impurity densities $n_Z {\,=\,} \unit[10^{19}]{m^{-3}}$ but higher deuterium densities, $n\sub{D} {\,=\,} \unit[10^{21}]{m^{-3}}$, the lower avalanche multiplication factor from figure~\ref{fig:mult1} results in a significantly lower maximal runaway current:  
\begin{equation*}
I\sub{RE} = \unit[2.3]{MA} \,({\rm Ne}), \quad I\sub{RE} = \unit[2.5]{MA}\, ({\rm Ar}),
\end{equation*}
which is only slightly higher than the completely screened result $I\sub{RE} {\,=\,} \unit[1.6]{MA}$ (for both argon and neon as $Z\sub{eff} \,{\approx}\,1$). 

{\it Discussion and conclusions.}---In this Letter, we present a calculation of the avalanche growth rate of runaway electrons in the presence of partially ionized impurities, summarized by the semi-analytic formula~\eqref{eq:GammaInterp}. 
We evaluate the avalanche amplification following a thermal quench induced by massive material injection, assuming a constant impurity density profile at fixed temperature. 
Our findings %, illustrated in figure~\ref{fig:mult1}, 
are striking: the injection of impurities can exacerbate the runaway avalanche problem, and with an ITER-like deuterium density and impurity densities near $10^{20}\,$m$^{-3}$, the avalanche amplification of a runaway seed can be in excess of $10^{35}$. This can be compared with the Rosenbluth--Putvinski estimate with values near $10^{16}$. Density regimes of such strongly enhanced runaway growth rate should be carefully avoided in ITER. 
%Higher temperatures during the current quench also help reduce the runaway avalanche generation.

The ITER disruption mitigation system employs high-$Z$ material injection in order to reduce the current quench time because of thermal and force load constraints. 
For runaway mitigation purposes, it has been suggested that this should be combined with massive deuterium injection in order to suppress runaway generation by raising the critical electric field without violating the current quench time requirements~\cite{Lehnen2015,MartinSolis2017}. Our results emphasize the importance of successfully raising the deuterium density throughout the post-disruptive ITER plasma, as this will effectively mitigate the detrimental effect of the impurities on runaway generation.
However, it is currently uncertain if it is feasible to quickly assimilate large amounts of material, and some experiments show poor penetration of injected material into the runaway-electron beam region~\cite{Reux2015,Nardon_2016}. Even provided that the combined injection of impurities and deuterium is successful, accurate models of avalanche generation in partially ionized plasmas are still needed in order to assess the efficacy of the method. 

%\clubpenalty=10000
%\widowpenalty=10000
%\brokenpenalty=5000
%\enlargethispage{\baselineskip}
For predictive modeling of disruption mitigation, it is necessary to include self-consistent impurity dynamics as well as radial runaway transport. 
%The avalanche growth rate obtained here can be readily implemented in disruption modeling tools, although this only allows for energy-independent radial transport. Energy-dependent transport coefficients can be included in an effective avalanche growth rate following the method in~\cite{HelanderDiffusion2000} with the updated growth rate \eqref{eq:GammaInterp}. 
%
The avalanche growth rate obtained here can be readily implemented in disruption modeling tools; with energy-independent transport, \eqref{eq:GammaInterp} can be directly used, whereas energy-dependent transport coefficients can be included in an effective avalanche growth rate following the method in~\cite{HelanderDiffusion2000} with the updated growth rate. 
%The avalanche growth rate obtained here can be readily implemented in fluid models with constant radial transport coefficients, whereas energy-dependent models can be included in an effective avalanche growth rate~\cite{HelanderDiffusion2000}. 
A simplified approach to runaway transport was suggested by Boozer~\cite{Boozer2019,
Boozer2015}. If the flux surfaces are broken during the disruption, and the runaways are lost before they reach relativistic speeds, it is only the poloidal flux change that occurs after flux surfaces have healed that contributes to runaway multiplication. In our model, $I_0$ and $n\sub{seed}$ should then denote the remaining plasma current and seed density once flux surfaces have healed. 

%Boozer~\cite{Boozer2019,Boozer2015} showed that if the magnetic flux surfaces are broken during the thermal quench, and the runaways are lost before they reach relativistic speeds, it is only the poloidal flux change that occurs after flux surfaces have healed that contributes to runaway multiplication. In our model, $I_0$ should then denote the remaining plasma current once flux surfaces have healed. Alternatively, at lower levels of transport, a local-diffusion model can be included as part of an effective avalanche growth rate~\cite{HelanderDiffusion2000}. 

%Finally, we note that present-day tokamaks are incapable of sustaining ITER-disruption-like electric fields for multiple avalanche times. Consequently, it is challenging to design experiments that directly validate the enhanced avalanche growth rates predicted by the model presented herein, which come into effect only at $E_\parallel \,{\gg}\,\Eceff$, and therefore depend on the collision frequencies $\bar\nu\sub{s}$ and $\bar\nu\sub{D}$ at low momenta. Support of the model may be obtained by comparing predictions of dissipation rates in the runaway plateau of today's high-$Z$ injection experiments, as these are set by \Eceff~\cite{breizman2014} which is sensitive to $\bar\nu\sub{s}$ and $\bar\nu\sub{D}$, albeit at high momenta.

\paragraph{Acknowledgments} The authors are grateful to G.~Papp, A.~Boozer, S.~Newton, R.~Mart\'in-Sol\'is, I.~Pusztai, M.~Hoppe and O.~Linder for fruitful discussions. This work was supported by the Swedish Research Council (Dnr. 2014-5510) and the European Research Council (ERC-2014-CoG grant 647121). 
\section*{References}
\bibliography{references_master}

\end{document}